\documentclass{aa}
\usepackage{graphicx}
\usepackage{txfonts}
\usepackage{natbib}
\usepackage{longtable}
\usepackage{hyperref} 
\usepackage{dsfont}
\hypersetup{colorlinks=true,linkcolor=blue,citecolor=blue,urlcolor=blue}
\usepackage{orcidlink}
\usepackage{lscape}
\usepackage{multirow}
\usepackage{multicol}

\begin{document}






\title{Yields from massive stars in binaries}
\subtitle{Chemical evolution of the Milky Way disk}

\author {E. Pepe \orcidlink{0009-0006-9760-3335} \inst{1} \thanks{email to: EMANUELE.PEPE@studenti.units.it},  
M. Palla \orcidlink{0000-0002-3574-9578}\inst{2,3}, 
F. Matteucci \orcidlink{0000-0001-7067-2302} \inst{1,4,5} \and
E. Spitoni \orcidlink{0000-0001-9715-5727}\inst{4, 6}  }
\institute{
   Dipartimento di Fisica, Sezione di Astronomia,
  Universit\`a di Trieste, Via G.~B. Tiepolo 11, I-34143 Trieste  \and 
  Dipartimento di Fisica e Astronomia “Augusto Righi”, Alma Mater Studiorum, Università di Bologna, Via Gobetti 93/2, I-40129 Bologna, Italy \and 
  INAF – Osservatorio di Astrofisica e Scienza dello Spazio di Bologna, Via Gobetti 93/3, I-40129 Bologna, Italy \\
              \email{marco.palla@inaf.it} \and 
  INAF - Osservatorio Astronomico di Trieste, via G.B. Tiepolo
 11, I-34131, Trieste, Italy\\
              \email{francesca.matteucci@inaf.it} \and  
 INFN - Sezione di Trieste, via Valerio 2, I-34134 Trieste, Italy
 \and IFPU Institute for Fundamental Physics of the Universe, Via Beirut 2, I-34151 Trieste, Italy
}

 \date{Received xxxx / Accepted xxxx}

\abstract {A large fraction of massive stars in the Galaxy reside in binary systems and their evolution  is different from that of single stars. The yields of massive stars, which are the main responsible for the production of metals, can be therefore affected by the binary nature of the systems. However, very few papers up to now have explored the effects of massive interacting binaries on the chemical evolution of the Milky Way. Recently, \citet{Farmer2023} computed new grids of yields for single and binary-stripped massive stars with solar chemical composition. The main purpose of this paper is to test these yields on the chemical evolution of Galactic stars. To do that, we adopt well-tested chemical evolution models for the Milky Way disk, implementing both yields for single and binary-stripped massive stars. In particular, we assume different percentages of massive binary systems within the initial mass function. We compute the evolution of 22 chemical species starting from $^{4}$He to $^{64}$Zn, including CNO, $\alpha$-elements and Fe-peak elements. Our main results can be summarized as follows: i) when adopting the yields of \citet{Farmer2023}, large differences are found relative to the predicted solar abundances by chemical evolution models adopting "standard" massive star yields from the literature for $^{12}$C,  $^{14}$N,  $^{24}$Mg,  $^{39}$K, $^{40}$Ca, $^{55}$Mn and $^{59}$Co.
Generally, the yields for single stars reproduce slightly better the observed solar abundances, although for several elements a large fraction of binaries helps in reproducing the observations; 
ii) different fractions of massive binaries (from 50\% to 100\%) produce negligible differences in the predicted solar abundances, whereas the differences are more marked between models with and without binary-stripped stellar yields; iii) for the [X/Fe] vs. [Fe/H] relations, the yields including massive stars in binaries produce the best results for $^{52}$Cr, while for $^{12}$C, $^{39}$K, $^{40}$Ca and $^{24}$Mg the best results are obtained with Farmer's yields with no binaries.
}


\titlerunning{Yields from massive binaries}

\authorrunning{E. Pepe et al.}

\maketitle


\section{Introduction}

Galactic archaeology deals with the interpretation of the observed chemical abundances in stars and gas in order to reconstruct the history of star formation and evolution  of our Galaxy and external ones. 
Among the different processes which are fundamental to compute the chemical evolution of galaxies, stellar yields, i.e. the amount of different chemical elements produced by stars and ejected into the interstellar medium (ISM), are one of the key ingredients to properly model the chemical enrichment in galaxies, but is also the most important source of intrinsic uncertainty within models (see, e.g. \citealt{romano2010}; \citealt{matteucci2021}, and references therein).

Detailed models of galactic chemical evolution include the yields of a large network of chemical elements starting from hydrogen up to the heaviest ones. Most of the metal mass is formed in massive stars ($M\gtrsim 10M_{\odot}$), which end their lives as core-collapse supernovae (CC-SNe). 
Generally, in models of galactic chemical evolution only the yields from single massive stars are considered (e.g. \citealt{WW1995}; \citealt{koba2006}; \citealt{nomoto2013}; \citealt{limongi2018}, but see \citealt{dedonder2007}), as large grids of stellar masses and elemental abundance yields are only available for single stellar models. 
However, stars preferentially form in clusters and associations (\citealt{lada2003}), where interactions are frequent. Therefore, binary stars are common, with binarity changing the stellar evolutionary paths, and the nucleosynthesis products relative to the case of single stars (see, e.g. \citealt{langer2012}; \citealt{woosley2019}).
In fact, during the evolution of a binary system several common envelope phases, including episodes of mass transfer/stripping in stars, can occur and change the element production, as the matter accreted/stripped is available/lost for subsequent nuclear processing.\\
To date, the only study focusing on the effect of binaries on the Milky Way (MW) chemical evolution is from \citet{dedonder2002}. In this study, the authors computed chemical yields from massive stars in binaries and tested them in a two-infall chemical model for the MW similar to that of \citet{chiappini1997}. 
Their results showed that including  massive binaries improved the agreement with the time evolution of carbon abundance and suggested that binary evolution can provide production of primary nitrogen from massive stars, although this can be also obtained from single rotating massive stars (e.g. \citealt{meynet2002}, \citealt{limongi2018}). 
On the other hand, they concluded that accounting for chemical enrichment of massive binaries produces a variation in the evolution of other chemical elements (He, O, Ne, Mg, Si, S, and Ca) of no more than a factor of 2 relative to the evolution without binaries.\\

Recently, \citet{Farmer2023} published yields for massive stars in binary systems for more chemical species than in the \citet{dedonder2002} and showed that differences in the production of elements in presence of binaries do exist for specific elements.
As for example, \citet{Farmer2023} showed that in these binary-stripped stars there is an increased production of F and K, relative to single stars. 

In this paper, we adopt a detailed and well-tested (e.g. \citealt{spitoni2019}, \citeyear{spitoni2020}; \citealt{palla2020}, \citealt{palla2021}) chemical evolution model for the MW, where the evolution of the abundances of several species (H, He, C, N, $\alpha$-elements, Fe and Fe-peak elements) is followed.
In particular, we introduce for the first time the  yields by \citet{Farmer2023} for single and binary-stripped stars in chemical evolution models for the MW and predict the solar chemical abundances as well as the [X/Fe] vs. [Fe/H]\footnote{[X/Y] = log(X/Y) - log(X$_\odot$/Y$_\odot$), where X, Y are the abundances of the object studied and X$_\odot$, Y$_\odot$ are solar abundances.} relations, where X are all the elements except Fe (which is the tracer of stellar metallicity). 
The results of the models are then compared with the abundance patterns as observed by large-scale surveys (APOGEE, \citealt{apogeedr172022}) as well as by smaller programs at higher spectral resolution (e.g. \citealt{bensby2014,nissen2020}), to test whether the yields adopted in this work can improve the agreement between the predicted abundance trends and observations in the MW disk.\\

The paper is therefore organized as follows: in Section \ref{models_sec} we describe the chemical evolution model, with a special focus on the description of the adopted stellar yields both for single and binary massive stars. In Section \ref{datasamp} we describe the observational data adopted for comparison with our results. In Section \ref{results_sec} we show and discuss the results of our predictions compared to observations, and in Section \ref{conclu_sec} we draw some conclusions.


\section{Chemical Evolution framework}
\label{models_sec}

In this Section, we present the chemical evolution models adopted throughout our work. 
The models used are as follows: 

\begin{enumerate}    
    \item the one-infall model (as proposed by \citealt{chiosi1980}; \citealt{matteucci1986}; \citealt{matteucci1989}; \citealt{boissier1999}), as described in \citet{matteucci2021}. 
    The one-infall model assumes that the Galactic disk components form sequentially as a result of a single infall episode of primordial gas, with a timescale $\tau\simeq 7$ Gyr for solar vicinity, in order to reproduce the G-dwarf metallicity distribution \citep{matteucci2012}.
    
    \item a revised two-infall model \citep[e.g.]{palla2020}, which assumes that the MW disk forms in two separate and sequential gas accretion episodes. The first one forms the chemical thick disk\footnote{here, for chemical thick and thin disk we refer to the high-$\alpha$ and low-$\alpha$ sequences observed in the MW disk.} with a timescale $\tau_1\simeq 1$ Gyr, while the second forms the chemical thin disk with a slower timescale $\tau_2\simeq 7$ Gyr.
    The term "revised" refers to the adoption of a larger delay of 3.25 Gyr between the first and second infall episode, at variance with the "classical" delay of 1 Gyr (\citealt{chiappini1997}; \citealt{romano2010}). The adopted assumptions in the "revised" two-infall model allowed us to reproduce large survey abundance data (\citealt{palla2020}; \citealt{spitoni2021}), as well as abundance-age diagrams (\citealt{spitoni2019}; \citealt{spitoni2020}) in the solar neighborhood.   
\end{enumerate}

In both models, the basic equation used to describe the evolution of an element $i$ in the ISM is (see \citealt{matteucci2021}):

\begin{equation}
    \label{}
    \dot{\sigma}_i(R,t)=-\psi(R,t)X_i(R,t)+\dot{R}_i(R,t)+\dot{\sigma}_{i,inf}(R,t).
\end{equation}

On the left-hand side, $\sigma_i(R,t)=\sigma_{gas}(R,t)X_i(R,t)$ is the fractional surface mass density of the element $i$ in the ISM at the time t, with $X_i(R,t)$ being the mass abundance of said element and $\sigma_{gas}(R,t)$ being the mass density of the ISM. The first term on the right-hand side represents the rate at which chemical elements are subtracted from the ISM by star formation, with $\psi(R,t)$ being the star-formation rate (SFR), parameterized according to the Schmidt-Kennicutt law \citep{kenni1998}:
\begin{equation}\label{}
    \psi(R,t)=\nu\, \sigma_{gas}^k(R,t),
\end{equation}
where $k=1.5$, and $\nu$ is the star formation efficiency expressed in units of Gyr$^{-1}$ and considered variable with the Galactocentric distance as in \citet{palla2020}. 

The second term of the equation, $\dot{R}_i(R,t)$, concerns the returned mass into the ISM in form of the new and old chemical element $i$. It represents the rate at which chemical elements are returned to the ISM through stellar winds and supernova explosions. The term $R_i(R,t)$ depends also on the initial mass function (IMF), here parameterized as in \citet{kroupa1993}.

The last term in the equation, $\dot{\sigma}_{i,inf}(R,t)$, is the gas infall rate,  which in the one-infall model is computed as:
\begin{equation}
\dot{\sigma}_{i,inf}(R,t)=A(R)\, X_{i,inf}\, e^{-\frac{t}{\tau}},   
\end{equation}
where $\tau$ is the infall timescale, $X_{i,inf}$ is the composition of the infalling gas (assumed to be primordial) and $A(R)$ is the normalizing factor that is chosen to reproduce the total surface mass density observed at the present-day at each radius (see also \citealt{palla2021}).

For the two-infall scenario, the gas infall rate is instead computed in this way:\\
\begin{equation}
    \dot{\sigma}_{i,inf}(R,t)=A(R)\, X_{i,inf}\, e^{-\frac{t}{\tau_1}} + \theta(t-t_{max})\, B(R)\, X_{i,inf}\, e^{-\frac{t-t_{max}}{\tau_2}}
\end{equation}
with $\tau_1$ and $\tau_2$ the infall timescales for first and second infall episode, $t_{max}$ the time of maximum infall which is also the delay between the two episodes, and $A(R)$ and $B(R)$ the coefficients obtained by reproducing present-day surface mass density of thick and thin disk in solar neighborhood. We also remind the reader that the $\theta$ in the equation above is respectively the Heavyside step function.\\

The model, besides the core-collapse SN rate, includes a detailed computation of the Type Ia SN rate assuming the single degenerate 
scenario and in particular the delay-time-distribution (DTD) function as computed by \citet{MR01}, which can be considered a good compromise to describe the delayed pollution from the entire SN Ia population (see \citealt{palla2021} and references therein). 

We note that neither of the above models assumes the presence of galactic winds.  \citet{melioli2008,melioli2009} and \citet{spitoni2008, spitoni2009}, while investigating Galactic fountains caused by Type II supernova (SNe) explosions in OB associations within the solar annulus, found that the metals ejected by these events fall back to nearly the same Galactocentric region from which they originated, thus having a little effect on the overall chemical evolution of the Galactic disk. Moreover, these findings were recently confirmed by \citet{hopkins2023}, who showed that the vast majority of the mass ejected from the disk is accreted  again on short timescales and near to the original ejection site (Galactic fountains). 

\subsection{Nucleosynthesis prescriptions}
\label{nucleo_subsec}

In this work, we adopt for the first time the stellar yields by \citet{Farmer2023} for massive stars in the context of well-tested models of chemical evolution for the MW. \cite{Farmer2023} estimate stellar yields for elements up to Zn for an extensive grid ($M_{ini}$ = 11 to 45 M$_\odot$) of both single and binary-stripped stars at solar metallicity using the MESA stellar evolution code (version 12115, see e.g. \citealt[\citeyear{paxton2013}, \citeyear{paxton2015}, \citeyear{paxton2018}, \citeyear{paxton2019}]{paxton2011}, \citealt{jermyn2023}). In \citet{Farmer2023}, stars are evolved from the zero-age Main Sequence to core-collapse and then supernova until shock breakout. The assumed stellar chemical composition in the models is the solar one (\citealt{Grevesse98}).
For binary stars,  \citet{Farmer2023} evolve the primary star considering a companion with mass ratio $M_2/M_1=0.8$ and initial orbital period between 38 and 300 days, in order to assure that all the binary stars undergo case B\footnote{the primary, evolving star in the binary system fill the Roche lobe for the first time after core-hydrogen exhaustion} mass transfer (\citealt{paczynski1967}; \citealt{vandenheuvel1969}). For these systems, the secondary star is modeled as a point-mass until the end of core-helium burning, at which point the secondary is removed and the primary star evolves until core-collapse (see \citealt{laplace2020}).
Therefore, the resulting yields refer only to the primary star in the binary system.
In both single and binary cases, stellar models are non-rotating.  
For more details about the choice of physics and model assumptions, we refer to \citet[\citeyear{Farmer2023}]{farmer2021} and \citet[\citeyear{laplace2021}]{laplace2020}.

In this study, we adopt different fractions of massive star binaries in the initial mass function to explore the effect of binary-stripped stars to chemical enrichment. In particular, we test the cases of 100\%, 70\%, 50\% and 0\%. The reason for passing from a binary fraction of 0\% to 50\% is that, if we assume percentages of binaries lower than 50\%, there are negligible differences in the results relative to the case with no binaries.\\

To make a comparison between the newly proposed yields by \citet{Farmer2023} with other well-tested yields for single  massive star from the literature, we also adopt the yield sets suggested in \citet[their model 15]{romano2010}. In particular, these yields consist in a combination of models obtained with the Geneva stellar evolutionary code (\citealt{meynet2002}; \citealt[\citeyear{hirschi2007}]{hirschi2005}; \citealt{ekstrom2008}) for elements lighter than O, and  those of \citet{koba2006} for heavier elements (see \citealt{romano2010} for more details).\\

Despite focusing on the outcome of different massive star yields, the model also includes the nucleosynthesis from low and intermediate mass stars (LIMS) and Type Ia SNe to properly account for the Galaxy chemical evolution.
In order to highlight the different enrichment from different massive star yields, we adopt for all the model of this paper the yields of \citet{karakas2010} for LIMS and the ones by \citet[their W7 model]{iwamoto1999} for Type Ia SNe.


\section{The data sample}
\label{datasamp}

In this study we use abundances of solar neighborhood stars as measured in APOGEE DR17 \citep{apogeedr172022}, \citet{bensby2014} and \citet{nissen2020}.
In the following, we provide further details on the different datasets adopted, also specifying which chemical elements are selected to perform our comparison.

\subsection{The APOGEE DR17 data sample}
\label{apogee}

Throughout this work, we adopt data from the high-resolution spectroscopic survey APOGEE DR17 \citep{apogeedr172022}, which is part of the Sloan Digital Sky Surveys (SDSS). APOGEE operates using the du Pont Telescope and the Sloan Foundation 2.5 m Telescope  \citep{gunn2006} at Apache Point Observatory. Stellar parameters and abundances are derived using the APOGEE Stellar Parameters and Chemical Abundance Pipeline (ASPCAP; \citealt{gperez2016}).  The model atmospheres used in APOGEE DR17 are based on the MARCS model \citep{gustafsson2008}, as discussed by \citet{jonsson2020}, and the line list is described in \citet{smith2021}.

Here, we consider only stars with a Galactocentric distance 7 ${\text kpc} \le R_{GC} \le 9$ kpc as computed in \citet{leung2019}   and reported in the value-added astroNN\footnote{\url{https://data.sdss.org/sas/dr17/apogee/vac/apogee-astronn/}} catalogue, where accurate distances for distant stars are obtained using a deep neural network trained on parallax measurements of nearby stars shared between Gaia (\citealt{gaia2016,Egaia2021}) and APOGEE.
Following the work of \citet{spitoni2024}, we also applied a further selection based on signal-to-noise ratio (SNR) and vertical height above/below the Galactic plane (|z|):  SNR>80 and |z|$\le$2 kpc. This selection leaves us with a sample of around 55111 total stars, with 55016 spectra observed for the C, 55042 for the O, 55047 for the Mg, 54765 for the K, 55015 for the Ca, 53445 for the Ti and 53267 for the Cr.

\subsection{Bensby et al. 2014 and Nissen et al. (2020) datasets}

In addition to the APOGEE data, in this work we consider abundance data from small programs also targeting disk stars (\citealt{bensby2014,nissen2020}), but in the optical wavelength range at very high spectral resolution ($R/R_{\rm APOGEE}>2$). In this way, we can perform further comparison with the predicted trends as we are probing a very different observational setup that may lead to differences in the derived abundances (e.g. \citealt{spina2022,Hegedus22}, see also later in the text).\\

In \citet{bensby2014}, chemical abundances were derived for 714 FG dwarf and subgiant stars in the solar neighborhood. Observations were conducted using various spectrographs (e.g., FIES, UVES, HARPS, MIKE) at multiple observational facilities (e.g., NOT, VLT, La Silla 3.6m, Magellan Clay). All observations were carried out at a resolution of $R > 40,000$, achieving high signal-to-noise ratios (S/N $> 150$). In this study, we adopt the stellar abundances of Fe, O, Mg, Ca, Ti, and Cr measured for all 714 stars in the original \citet{bensby2014} sample.

The abundances from \citet{nissen2020}, on the other hand, were derived from very high-resolution (R $> 100,000$), high signal-to-noise (S/N $> 600$) observations of 72 solar twin stars, obtained using the HARPS and HARPS-N spectrographs at the La Silla 3.6m and TNG telescopes. These data provide highly precise measurements of elemental abundances in the solar vicinity. For this study, we adopt the stellar abundances of Fe, C, O, Mg, Ca, Ti, and Cr for all stars in the \citet{nissen2020} sample.

\begin{table}[]
 \caption{List of models adopted in this work.}
\resizebox{\columnwidth}{!}{%
\begin{tabular}{cccc}
\hline
Model          & Stellar Yields      & \% of binaries &     Evolutionary scenario \\[0.03cm]
\hline
K0-1       & see \citet{romano2010}     & -              & One-infall\\[0.04cm]
K0-2       & see \citet{romano2010}     & -              & Two-infall\\[0.05cm]
F0-1       & \citet{Farmer2023}         & 0\%            & One-infall\\[0.04cm]
F0-2       & \citet{Farmer2023}         & 0\%            &  Two-infall\\[0.05cm]
F50-1      & \citet{Farmer2023}         & 50\%           &  One-infall\\[0.04cm]
F50-2      & \citet{Farmer2023}         & 50\%           &  Two-infall\\[0.05cm]
F70-1      & \citet{Farmer2023}         & 70\%           &  One-infall\\[0.04cm]
F70-2      & \citet{Farmer2023}         & 70\%           &  Two-infall\\[0.05cm]
F100-1     & \citet{Farmer2023}         & 100\%          &  One-infall\\[0.04cm]
F100-2     & \citet{Farmer2023}         & 100\%          & Two-infall\\ 
\hline
\end{tabular}%
}
\label{model}
\end{table}


\section{Results}
\label{results_sec}

In this Section, we  show the results obtained by our chemical evolution models testing different setups for CC-SN yields, including models of stars in binary systems. We have chosen to show two basically different kind of models: one-infall and two-infall ones. The reason is that, although the one-infall model is only good for describing the evolution of the thin disk, it shows more clear predicted behaviors of the [X/Fe] vs. [Fe/H] relations, and it allows an easier comparison between cases with different stellar yields. On the other hand, the two-infall model more realistically describes the evolution of the thick and thin disk.

The models and their prescriptions are reported in Table \ref{model}, where we show the model name in the first column, the adopted massive star yields in the second column, the percentage of binaries considered in the case of adoption of the \citet{Farmer2023} yields in the third column, and whether we adopt a one-infall or two-infall scheme for chemical evolution. 
Throughout the rest of the paper, we will refer to the models labeled with "K0" as Reference Models, as they adopt literature well-tested yield sets for single massive stars as described in Section \ref{nucleo_subsec} (see \citealt{romano2010}).
All the other models refer to the yields of \citet{Farmer2023} for massive stars, while the yields for other stellar types are the same as in K0 Models (see Section \ref{nucleo_subsec}). The numbers 1 or 2, in the label of the models, refer to the one- and two-infall model, respectively. It is worth noting that the Reference Models K0 contain exactly the same physical assumptions as the other models, as described in Section \ref{models_sec} except for the yields from massive stars, which are those adopted in \citet{romano2010}, as described in Section \ref{nucleo_subsec}.

\subsection{Solar abundances}
\label{solabund_subsec}

\begin{figure*}
    \centering
    \includegraphics[scale=0.38]{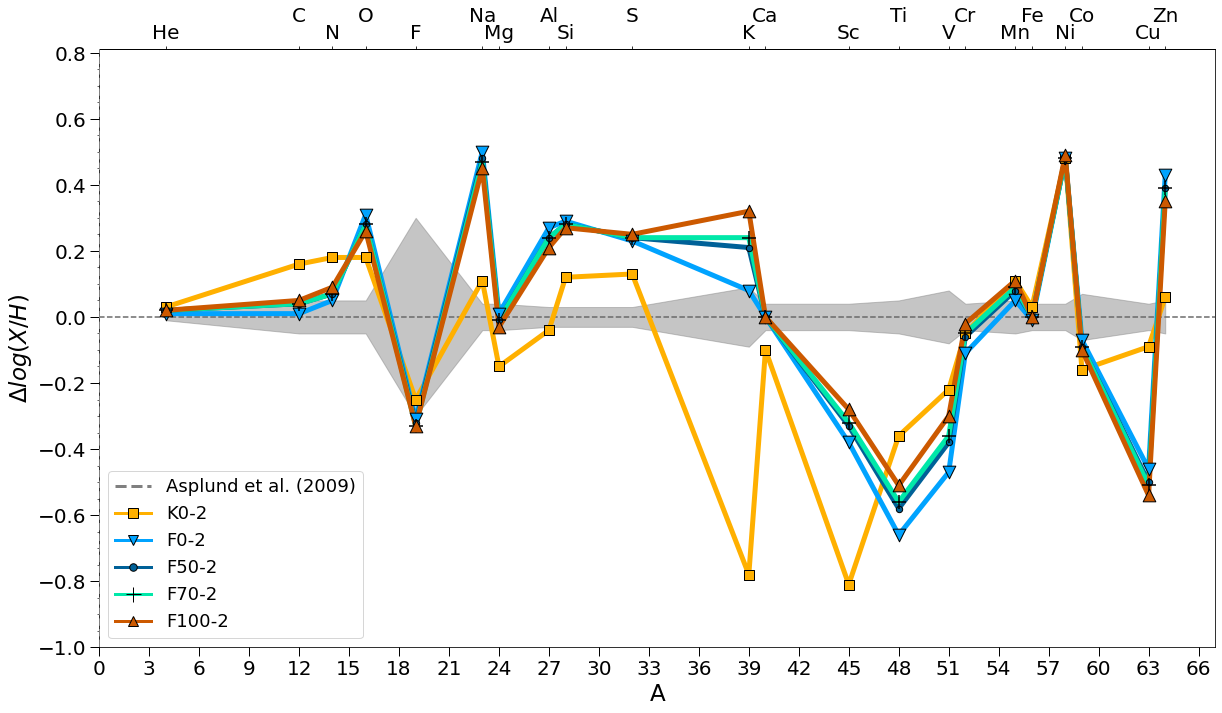}
    \caption{$\Delta log(X/H)$ ratios for different elements as predicted by the two-infall model for different stellar yields and percentage of binaries (see legend). Thin, gray dashed line indicates the solar ratios as measured in \citet{asplund09}, while gray shaded region is the abundance uncertainty.}
    
    \label{delta_abund_2inf}
\end{figure*}

In this first part, we show the model predictions obtained for the solar abundances, namely the ISM abundances at the time of birth of our Sun. Model abundances are taken at an age of $4.5$ Gyr ago, i.e. the time at which the proto-solar cloud was formed, and compared with measured solar abundances as obtained by \citet{asplund09}.\\

In Table \ref{abund_2inf} we show the the solar abundances as predicted by the two-infall model.  In particular, the abundances by number ($12+\log({\rm X}/{\rm H})$) of 22 elements from He to Zn are reported. As explained in Section \ref{models_sec}, the adopted chemical evolution model, has already been tested in several studies and allows us to best reproduce the multiple constraints coming from abundance patterns and ages in the solar vicinity (e.g. \citealt{spitoni2019,spitoni2020,palla2020}, \citealt{molero2023}).
Therefore, the model outputs can be used for an insightful comparison with the measured solar abundances. In column 1 of Table \ref{abund_2inf}
 we report the chemical species, in column 2 the observed solar abundances, then in column 3, 4, 5 and 6 the predictions from the models.\\
 
By looking at the Table, we can observe that once \citet{Farmer2023} stellar yields are used, different percentages of binary stars result in negligible variations for most of the chemical elements. The only notable differences are observed for $^{39}$K, $^{48}$Ti  and $^{51}$V.
On the other hand, more marked variations are seen between the results obtained with the yields of \citet{Farmer2023} and those used in \citet[i.e. our Reference Model, K0-2]{romano2010}.
In particular, the Reference Model is better in reproducing the solar abundances of $^{19}$F, $^{23}$Na, $^{27}$Al, $^{28}$Si, $^{32}$S, $^{51}$V, $^{63}$Cu and $^{64}$Zn. On the other hand $^{12}$C, $^{24}$Mg, $^{39}$K and $^{40}$Ca solar abundances are much better reproduced by models using \citet{Farmer2023} stellar yields. For $^{39}$K instead, the yields of \citet{Farmer2023} well reproduce the solar abundance only in the case where no binaries are assumed (Model F0-2). 
However, it should be noted that while the yields of \citet{romano2010} are metallicity dependent, those of \citet{Farmer2023} are computed only for the solar chemical composition, thus making such a comparison more difficult (see later Section \ref{caveats}).

This is even more evident when we observe Figure \ref{delta_abund_2inf}, where we show, for each element, the difference between the model predictions and the data from \citet{asplund09}, as reported in Table \ref{abund_2inf}.  The chemical elements in this figure are identified by the atomic mass number (A) of their main isotopes.

\begin{table}[]
\caption{Solar abundances as predicted by the two-infall model for the different yields tested in this work (see Table \ref{model}).}

\resizebox{\columnwidth}{!}{%
\begin{tabular}{|l|r|lllll|}
\hline
\multicolumn{1}{|c|}{\multirow{2}{*}{Element}} & \multicolumn{1}{c|}{\multirow{2}{*}{Asplund}} & \multicolumn{5}{c|}{Model}                                                                                                                      \\[0.03cm] \cline{3-7}
\multicolumn{1}{|c|}{}                          & \multicolumn{1}{c|}{}                         & \multicolumn{1}{r|}{K0-2}  & \multicolumn{1}{r|}{F0-2}  & \multicolumn{1}{r|}{F50-2} & \multicolumn{1}{r|}{F70-2} & \multicolumn{1}{r|}{F100-2} \\[0.03cm] \hline
$^{4}$He                                        & 10.93$\pm$0.01                                & \multicolumn{1}{l|}{10.96} & \multicolumn{1}{l|}{10.94} & \multicolumn{1}{l|}{10.95} & \multicolumn{1}{l|}{10.95} & 10.95                       \\[0.03cm] \hline
$^{12}$C                                        & 8.43$\pm$0.05                                 & \multicolumn{1}{l|}{8.59}  & \multicolumn{1}{l|}{8.44}  & \multicolumn{1}{l|}{8.47}  & \multicolumn{1}{l|}{8.47}  & 8.48                        \\[0.03cm] \hline
$^{14}$N                                        & 7.83$\pm$0.05                                 & \multicolumn{1}{l|}{8.01}  & \multicolumn{1}{l|}{7.88}  & \multicolumn{1}{l|}{7.90}  & \multicolumn{1}{l|}{7.90}  & 7.92                        \\[0.03cm] \hline
$^{16}$O                                        & 8.69$\pm$0.05                                 & \multicolumn{1}{l|}{8.87}  & \multicolumn{1}{l|}{9.00}  & \multicolumn{1}{l|}{8.97}  & \multicolumn{1}{l|}{8.97}  & 8.95                        \\[0.03cm] \hline
$^{19}$F                                        & 4.56$\pm$0.30                                 & \multicolumn{1}{l|}{4.31}  & \multicolumn{1}{l|}{4.25}  & \multicolumn{1}{l|}{4.24}  & \multicolumn{1}{l|}{4.23}  & 4.23                        \\[0.03cm] \hline
$^{23}$Na                                       & 6.24$\pm$0.04                                 & \multicolumn{1}{l|}{6.35}  & \multicolumn{1}{l|}{6.74}  & \multicolumn{1}{l|}{6.72}  & \multicolumn{1}{l|}{6.71}  & 6.69                        \\[0.03cm] \hline
$^{24}$Mg                                       & 7.60$\pm$0.04                                 & \multicolumn{1}{l|}{7.45}  & \multicolumn{1}{l|}{7.61}  & \multicolumn{1}{l|}{7.59}  & \multicolumn{1}{l|}{7.59}  & 7.57                        \\[0.03cm] \hline
$^{27}$Al                                       & 6.45$\pm$0.03                                 & \multicolumn{1}{l|}{6.41}  & \multicolumn{1}{l|}{6.72}  & \multicolumn{1}{l|}{6.69}  & \multicolumn{1}{l|}{6.69}  & 6.66                        \\[0.03cm] \hline
$^{28}$Si                                       & 7.51$\pm$0.03                                 & \multicolumn{1}{l|}{7.63}  & \multicolumn{1}{l|}{7.80}  & \multicolumn{1}{l|}{7.79}  & \multicolumn{1}{l|}{7.79}  & 7.78                        \\[0.03cm] \hline
$^{32}$S                                        & 7.12$\pm$0.03                                 & \multicolumn{1}{l|}{7.25}  & \multicolumn{1}{l|}{7.35}  & \multicolumn{1}{l|}{7.36}  & \multicolumn{1}{l|}{7.36}  & 7.37                        \\[0.03cm] \hline
$^{39}$K                                        & 5.03$\pm$0.09                                 & \multicolumn{1}{l|}{4.25}  & \multicolumn{1}{l|}{5.11}  & \multicolumn{1}{l|}{5.24}  & \multicolumn{1}{l|}{5.27}  & 5.35                        \\[0.03cm] \hline
$^{40}$Ca                                       & 6.34$\pm$0.04                                 & \multicolumn{1}{l|}{6.24}  & \multicolumn{1}{l|}{6.34}  & \multicolumn{1}{l|}{6.34}  & \multicolumn{1}{l|}{6.34}  & 6.34                        \\[0.03cm] \hline
$^{45}$Sc                                       & 3.15$\pm$0.04                                 & \multicolumn{1}{l|}{2.34}  & \multicolumn{1}{l|}{2.77}  & \multicolumn{1}{l|}{2.82}  & \multicolumn{1}{l|}{2.83}  & 2.87                        \\[0.03cm] \hline
$^{48}$Ti                                       & 4.95$\pm$0.05                                 & \multicolumn{1}{l|}{4.59}  & \multicolumn{1}{l|}{4.29}  & \multicolumn{1}{l|}{4.37}  & \multicolumn{1}{l|}{4.39}  & 4.44                        \\[0.03cm] \hline
$^{51}$V                                        & 3.93$\pm$0.08                                 & \multicolumn{1}{l|}{3.71}  & \multicolumn{1}{l|}{3.46}  & \multicolumn{1}{l|}{3.55}  & \multicolumn{1}{l|}{3.57}  & 3.63                        \\[0.03cm] \hline
$^{52}$Cr                                       & 5.64$\pm$0.04                                 & \multicolumn{1}{l|}{5.69}  & \multicolumn{1}{l|}{5.53}  & \multicolumn{1}{l|}{5.58}  & \multicolumn{1}{l|}{5.59}  & 5.62                        \\[0.03cm] \hline
$^{55}$Mn                                       & 5.43$\pm$0.05                                 & \multicolumn{1}{l|}{5.54}  & \multicolumn{1}{l|}{5.48}  & \multicolumn{1}{l|}{5.51}  & \multicolumn{1}{l|}{5.52}  & 5.54                        \\[0.03cm] \hline
$^{56}$Fe                                       & 7.5$\pm$0.04                                  & \multicolumn{1}{l|}{7.53}  & \multicolumn{1}{l|}{7.49}  & \multicolumn{1}{l|}{7.50}  & \multicolumn{1}{l|}{7.50}  & 7.50                        \\[0.03cm] \hline
$^{58}$Ni                                       & 6.22$\pm$0.04                                 & \multicolumn{1}{l|}{6.70}  & \multicolumn{1}{l|}{6.70}  & \multicolumn{1}{l|}{6.70}  & \multicolumn{1}{l|}{6.70}  & 6.71                        \\[0.03cm] \hline
$^{59}$Co                                       & 4.99$\pm$0.07                                 & \multicolumn{1}{l|}{4.83}  & \multicolumn{1}{l|}{4.92}  & \multicolumn{1}{l|}{4.90}  & \multicolumn{1}{l|}{4.90}  & 4.89                        \\[0.03cm] \hline
$^{63}$Cu                                       & 4.19$\pm$0.04                                 & \multicolumn{1}{l|}{4.10}  & \multicolumn{1}{l|}{3.73}  & \multicolumn{1}{l|}{3.69}  & \multicolumn{1}{l|}{3.68}  & 3.65                        \\[0.03cm] \hline
$^{64}$Zn                                       & 4.56$\pm$0.05                                 & \multicolumn{1}{l|}{4.62}  & \multicolumn{1}{l|}{4.99}  & \multicolumn{1}{l|}{4.95}  & \multicolumn{1}{l|}{4.95}  & 4.91                        \\[0.03cm] \hline
\end{tabular}%
}\vspace{0.2cm}
{\footnotesize {\bf Notes:} in the second column, solar abundance measurements from \citet{asplund09} are shown. Model predictions are obtained by considering the outputs 4.5 Gyr ago from the present-day.}
\label{abund_2inf}

\end{table}

\subsection{Chemical abundance patterns}
\label{solabund_subsec}

In this Section we present the results of [X/Fe] vs. [Fe/H] abundance ratios for our adopted models testing different stellar yield prescriptions.
It is worth noting that during this Section we display the results for both the one-infall and the two-infall scenarios.
In fact, we use the one-infall scenario to compare the abundance evolution predicted by models between each other, while the two-infall scenario is used to compare the chemical evolution models with the data as described in Section \ref{datasamp}. 
Such a choice is justified by the fact that the outputs of the one-infall scenario allows us to highlight and explain better the effects produced by the different yields for massive stars on the predicted abundance patterns, whereas the more physically robust star formation history of two-infall model is better suited to reproduce the observed abundance trends in the solar neighborhood.\\

In the following, we show the model results for the chemical elements that are more relevant for our study. 
In particular, we focus on the chemical abundances for which we have a large amount of data and we observe important differences in the evolution of the [X/Fe] vs. [Fe/H] abundance patterns for different yield prescriptions. In the [X/Fe] plots we will use always the same color system for the same four Models, as already used in Figure \ref{delta_abund_2inf}.
Moreover, as \citet{Farmer2023} yields are only computed for the solar metallicity, we also excluded from the analysis those elements which are known to show a marked dependence on the metallicity, namely the elements with a prominent secondary\footnote{the production of an element is primary when stems directly from the of H and He. Conversely, a secondary production imply metal seeds already present at stellar birth.} component (see later Section \ref{caveats}).

\subsubsection{$\alpha$-elements}
\label{alpha_subsubsec}

We start our analysis from $^{12}$C and the most relevant $\alpha$-elements, namely $^{16}$O, $^{24}$Mg, $^{40}$Ca and $^{48}$Ti.
It is worth noting that  all the model outputs and stellar data are normalized to \citet{asplund09} solar abundances, in agreement with the data presented in Section \ref{datasamp}.\\

It is worth reminding that the following [X/Fe] vs. [Fe/H] diagrams have to be interpreted according to the time-delay model (\citealt{tinsley1980, greggio1983, matteucci1986}; \citealt[\citeyear{matteucci2021}]{matteucci2012}). The time-delay interpretation stands on the fact that the [Fe/H]-axis can be interpreted as a time evolution axis. Therefore,  at low metallicities (hence at earlier times) there is a predominant contribution to metals from massive stars and only at larger metallicities there is a substantial production of Fe from Type Ia SNe, which start exploding with a delay relative to CC-SNe and can have explosion times as long as the Hubble time. 
According to this, the [$\alpha$/Fe] ratios show the so-called plateau at low metallicities ([Fe/H] $\lesssim-1.0$ dex in the MW), and then decline for larger metallicities as $\alpha$-elements should be predominantly produced by CC-SNe.  
If the element considered is instead more importantly produced by Type Ia SNe and/or by low and intermediate mass stars, the change in slope at intermediate-high metallicity is less marked and it becomes null or even positive when the element is produced mostly by delayed sources.

We note that the behaviors of the above elements relative to Fe, is different for the one-infall and two-infall model. In fact, in the latter, which is best suited to reproduce the multiple features observed in the MW disk, there is a natural gap in the SFR between the formation of the chemical thick and thin disks. This gap produces the loops observed in the bottom panels of the following figures.
This behavior is the consequence of a delayed second gas infall, which dilutes the ISM with primordial gas, lowering the [Fe/H] ratio and leaving the [X/Fe] unchanged. The metal abundance is then restored thanks to the subsequent episode of star formation (see also \citealt{spitoni2019}).\\

\begin{figure}
    \centering
    \includegraphics[scale=0.36]{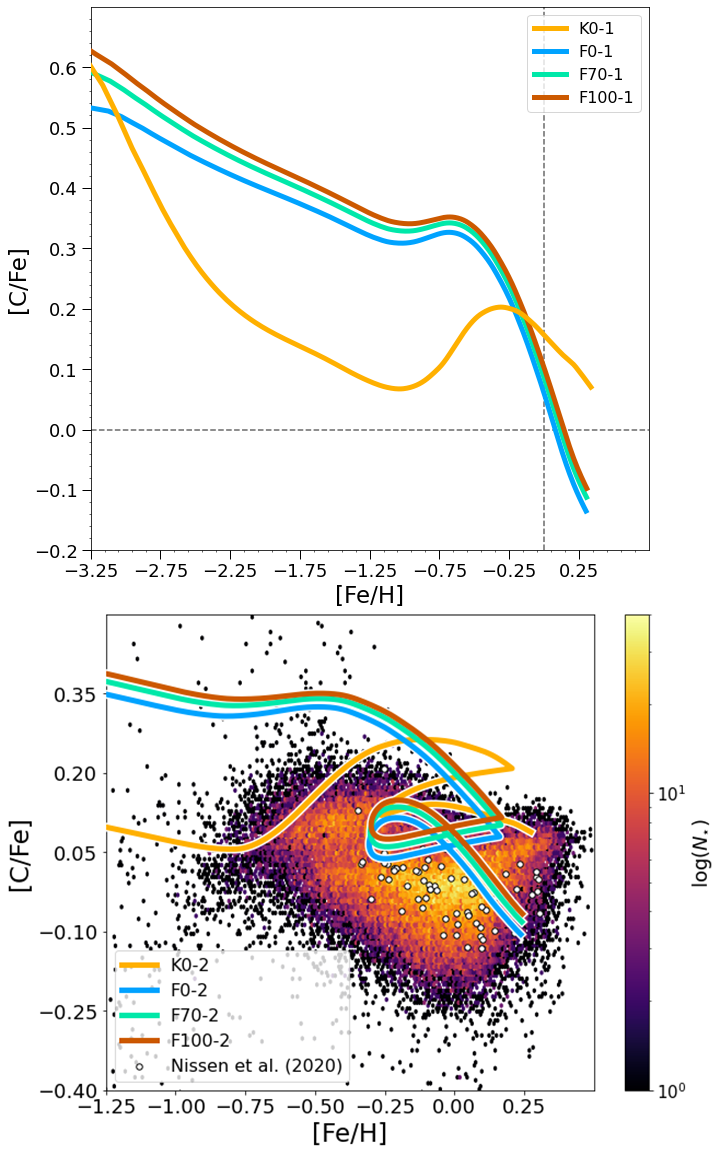}
    \caption{Upper panel: [C/Fe] vs. [Fe/H] ratios as predicted by the one-infall model for different stellar yields and percentage of binaries (see legend). Thin, gray dashed lines indicate the solar ratios. Lower panel: [C/Fe] vs. [Fe/H] ratios as predicted by the two-infall model for different stellar yields and compared with data from  \citet[white points]{nissen2020} and APOGEE \citep{apogeedr172022}. }
    
    \label{C_map}
\end{figure}

Going into the details of chemical elements, in Figure \ref{C_map}, we compare results from our models for [C/Fe] vs. [Fe/H].
In the upper panel, we compare the one-infall model (with the Reference Model being K0-1) for different percentages of binaries, namely 0\%, 70\% and 100\%, in the IMF of massive stars.
Models that use \citet{Farmer2023} yields show an almost typical $\alpha$ behavior, with a plateau/slow decrease in the [C/Fe] ratio at low metallicity, a knee around [Fe/H] $\sim-1$ dex  and a steeper decrease at high metallicity.  Moreover, we observe that models with higher binary fractions tend to rise the level of the plateau, highlighting that yields from binary massive stars predict a larger C enrichment (see \citealt{farmer2021}, \citealt{romano2022}).
The same $\alpha$-element behavior is instead not observed in the Reference Model, i.e. with same yield prescription as the model 15 in \citet{romano2010}: for this model, we observe a low-metallicity steep decrease in [C/Fe]  followed by an increase (at [Fe/H]$\simeq -1$ dex) and again a decrease in this ratio. The lower [C/Fe] ratios predicted by the Reference Model are due to the lower C yields  (\citealt{meynet2002}; \citealt{hirschi2007}; \citealt{ekstrom2008}) at low metallicities, with only a significant contribution by extremely metal-poor massive stellar rotators (EMP, [Fe/H] $<-3$ dex).
Due to the lower ratio at metallicities around [Fe/H] $\sim-1$ dex, the Reference Model shows a prominent bump due to the contribution of low-mass AGB stars to the carbon enrichment (see also \citealt{romano2019,ventura22}), which is instead almost hidden in the models adopting \citet{Farmer2023} yields. 
However, the comparison shows that the larger C yields in metal-rich SNe in the Reference Model prevent the steep decrease in [C/Fe] and produce a too large ratio at solar metallicity, at variance with what happens using \citet{Farmer2023} yields.

In Fig. \ref{C_map} lower panel, we compare results by the K0-2 (Reference Model), F0-2, F70-2 and F100-2 Models for [C/Fe] vs. [Fe/H] with data from stars in the solar vicinity, as described in Section \ref{datasamp}.
All the models displayed in the lower panel show some difficulties at reproducing the trends shown by the data.
In particular, the Reference Model K0-2 severely overestimates the data from \citet{nissen2020} and APOGEE (\citealt{apogeedr172022})  at [Fe/H] $>-0.5$ dex, while it aligns with the trend observed at lower metallicities (see also \citealt{romano2010}; \citealt{romano2019}).  
On the other hand, the models using \citet{Farmer2023} yields overestimate the observed [C/Fe] ratio at low metallicities, although they are in relatively good agreement with the sample of \citet{nissen2020} and APOGEE \citep{apogeedr172022} data at solar and super-solar metallicities. Therefore, the massive star yields from \citet{Farmer2023} are better tracers of the C enrichment at high metallicities, whereas those from \citet{koba2006} are in better line with the trends observed at lower metallicities.\\ 

\begin{figure}
    \centering
    \includegraphics[scale=0.39]{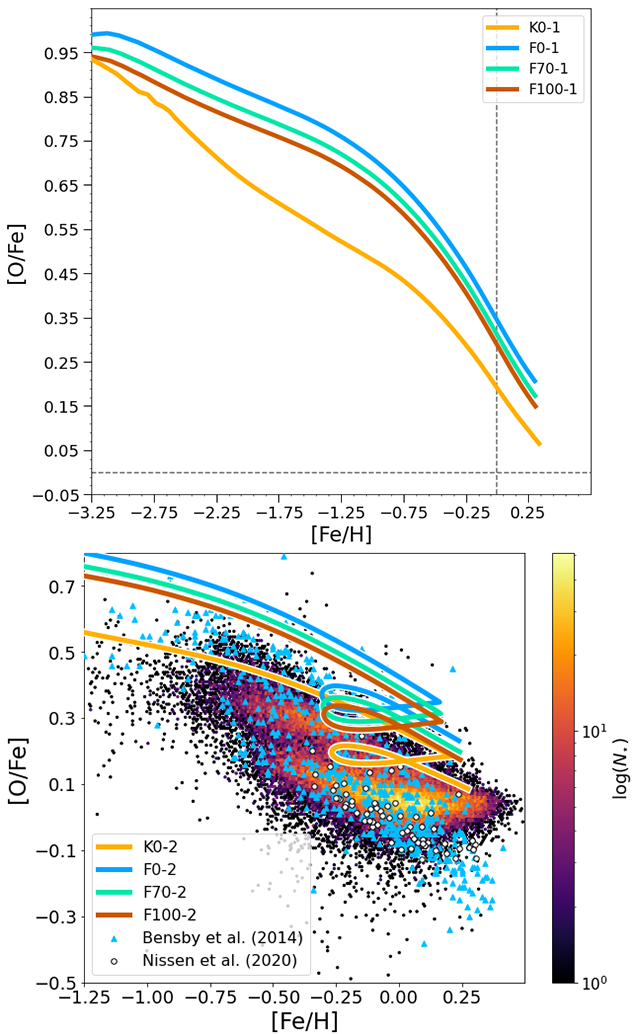}
    \caption{Same as Fig. \ref{C_map} but for [O/Fe]. Data are from \citet[azure triangles]{bensby2014}, \citet[white points]{nissen2020} and APOGEE \citep[colored according to their number density, see colorbar]{apogeedr172022}.}
    \label{O_map}
\end{figure}

In Figure \ref{O_map}, we compare results from our models for [O/Fe] vs. [Fe/H]. Upper and lower panels show the results of the same models as in Figure \ref{C_map}, and 
all the subsequent figures will follow the same scheme.  
All the models presented in Figure \ref{O_map} upper panel show the characteristic $\alpha$-element behavior with a plateau/shallow slope at low metallicity and a steeper slope at higher metallicity, with a knee around [Fe/H] $\sim-1$ dex. This is the typical behavior of the "time-delay model" (see \citealt {tinsley1980}; \citealt{greggio1983}; \citealt{matteucci1986}; \citealt{chiappini1997}; \citealt[\citeyear{matteucci2021}]{matteucci2012}).
The models adopting \citet{Farmer2023} yields follow the same trend. 
However, it is worth noting that models assuming higher binary fractions show lower [O/Fe] ratio, at variance with  what happens for C.  This is due to the fact that a higher production of C from massive stars, necessarily results into a decreased O production, because the higher C produced and ejected through stellar winds (see \citealt{farmer2021,Farmer2023}) has been subtracted from being further processed into O.
For what concerns the Reference Model instead, the predicted [O/Fe] ratios have similar values relative to the models adopting \citet{Farmer2023} yields at low metallicities, and slightly lower values for [Fe/H] $>-2.25$ dex, evidencing lower O yields especially for massive stars with $m<20$ M$_{\odot}$.

In any case, all the models shown in Fig. \ref{O_map} upper panel show super-solar [O/Fe] values at all metallicities. This is reflected in the comparison with the solar vicinity data in the lower panel, where the models generally overestimate the observed [O/Fe] trends in different stellar samples.
In particular, the Model F0-2 (using \citealt{Farmer2023} yields assuming single stars only) does not match any of the survey data, while the other models (in particular the Reference Model K0-2) better reproduce the observed trends at low metallicity and especially the data by \citet{bensby2014}.
Nonetheless, it remains evident the overestimation of the [O/Fe] ratio by the different chemical evolutionary tracks around the solar metallicity, thus suggesting a significantly lower O ejection by massive stars relative to what predicted by the different stellar models.\\

\begin{figure}
    \centering
    \includegraphics[scale=0.35]{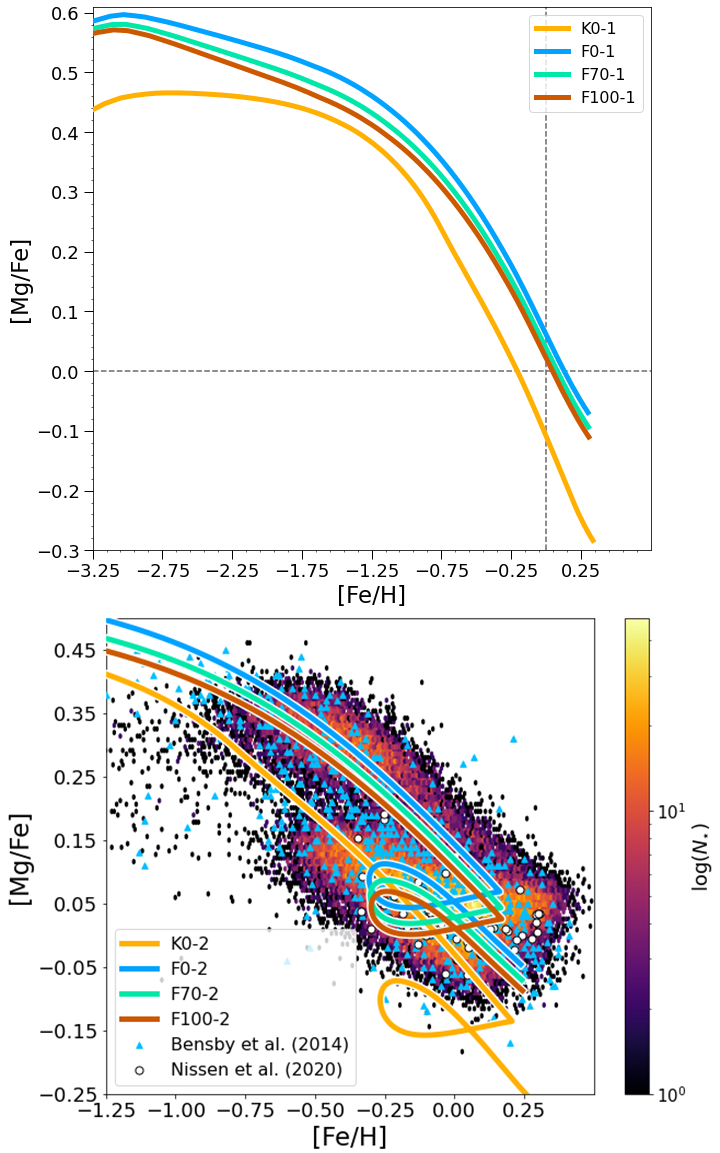}
    \caption{Same as Fig. \ref{C_map} but for [Mg/Fe]. Data are from \citet[azure triangles]{bensby2014}, \citet[white points]{nissen2020} and APOGEE \citep[colored according to their number density, see colorbar]{apogeedr172022}.}
    \label{Mg_map}
\end{figure}

Concerning Mg, we show our results in Figure \ref{Mg_map}.
The Reference Model adopting \citet{koba2006} yields shows the typical $\alpha$-element trend, with a plateau at low metallicity and a steep decrease in the [Mg/Fe] ratio after [Fe/H] $\sim-1$ dex. All models using yields from \citet{Farmer2023} show a similar pattern, typical of $\alpha$-elements. However, \citet{Farmer2023} models generally predict a higher value of the [Mg/Fe] ratio at at all metallicities, especially when only single star yields are employed, relative to the predictions by the Reference Model K0-2.
It is worth noting that the differences in the yields of Mg for single massive stars of \citet{Farmer2023} and those of \citet{koba2006} are mainly due to the assumption of overshooting during C burning in \citet{Farmer2023} models and not in the models by \citet{koba2006} (see also \citealt{Tominaga07}). Such an assumption can have a significant impact on the C yields but not only, as the $^{12}$C yield is sensitive to the size of the pocket of C that survives, which would then propagate also on the products of carbon burning, namely $^{20}$Ne,  $^{23}$Na, and $^{24}$Mg, and to successive burnings leading to heavier elements (see, e.g. \citealt[\citeyear{Farmer2023}]{farmer2021}).

Coming back to the difference observed in the chemical patterns of Fig. \ref{Mg_map} upper panel, such a difference is also evident in Figure \ref{Mg_map} lower panel:
in fact, the Reference Model shows a lack of agreement with the observational data from APOGEE \citep{apogeedr172022}, \citet{bensby2014} and \citet{nissen2020}, with a clear underestimation of the observed trend especially at high metallicities (see also \citealt{palla2022} for a comparison with a different sample of survey data).
On the other side, the two-infall models adopting different \citet{Farmer2023} yield sets show a  better agreement with all the data adopted in this study.
Focusing in particular on the APOGEE \citep{apogeedr172022} survey, we note that the F0-2 Model (assuming no binary-stripped star models) shows a remarkable agreement with both the observed high-$\alpha$ and the low-$\alpha$ sequences. This agreement slightly worsen when we increase the binary fraction in the models. However, the chemical tracks are still consistent with observations from the different data samples.\\

\begin{figure}
    \centering
    \includegraphics[scale=0.397]{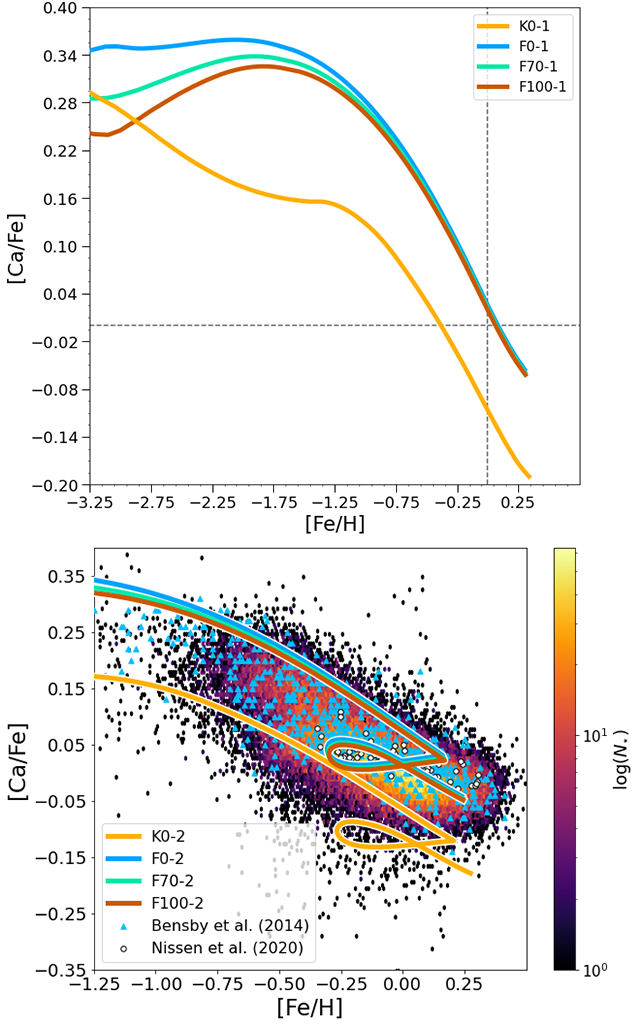}
    \caption{Same as Fig. \ref{C_map} but for [Ca/Fe]. Data are from \citet[azure triangles]{bensby2014}, \citet[white points]{nissen2020} and \citep[colored according to their number density, see colorbar]{apogeedr172022}.}
    \label{Ca_map}
\end{figure}

Another $\alpha$-element with quite significant variations between the different studied yield sets is Ca. 
In Figure \ref{Ca_map}, we compare results from our models for [Ca/Fe] vs. [Fe/H]. 
The upper panel highlights a different behavior either between \citet{koba2006} and \citet{Farmer2023} yield sets or between binary-stripped and single star models from \citet{Farmer2023}.
In fact, the model implementing \citet{koba2006} shows a lower [Ca/Fe] enrichment before and after the knee caused by the Type Ia SNe Fe production, with sub-solar [Ca/Fe] values at high metallicities and an offset of the order of $\sim 0.15$ dex relative to \citet{Farmer2023} yield sets.
For different \citet{Farmer2023} models, instead, we see that while single star yields produce a very flat plateau already at [Fe/H] $\lesssim-3$ dex, models with increasing binary fractions show lower values at such low metallicities.
This implies a significantly lower contribution to Ca enrichment by high-mass ($m\gtrsim20$ M$_\odot$, i.e. the first enriching with metals the ISM), binary-stripped stars, and a higher one in binary-stripped massive stars with lower masses ($m\lesssim20$ M$_\odot$).

In Figure \ref{Ca_map} lower panel, we can observe that the Reference Model does not agree with the observational data as it clearly underproduces [Ca/Fe] at all metallicities (see also \citealt{romano2010}).
Conversely, the models adopting \citet{Farmer2023} yields very well reproduce all the data. As also observed in Figure \ref{Ca_map} upper panel, there is no visible difference between the models at metallicity above [Fe/H] $\sim-1$ dex, despite varying the percentage of binaries inside the chemical evolution model.\\

\begin{figure}
    \centering
    \includegraphics[scale=0.39]{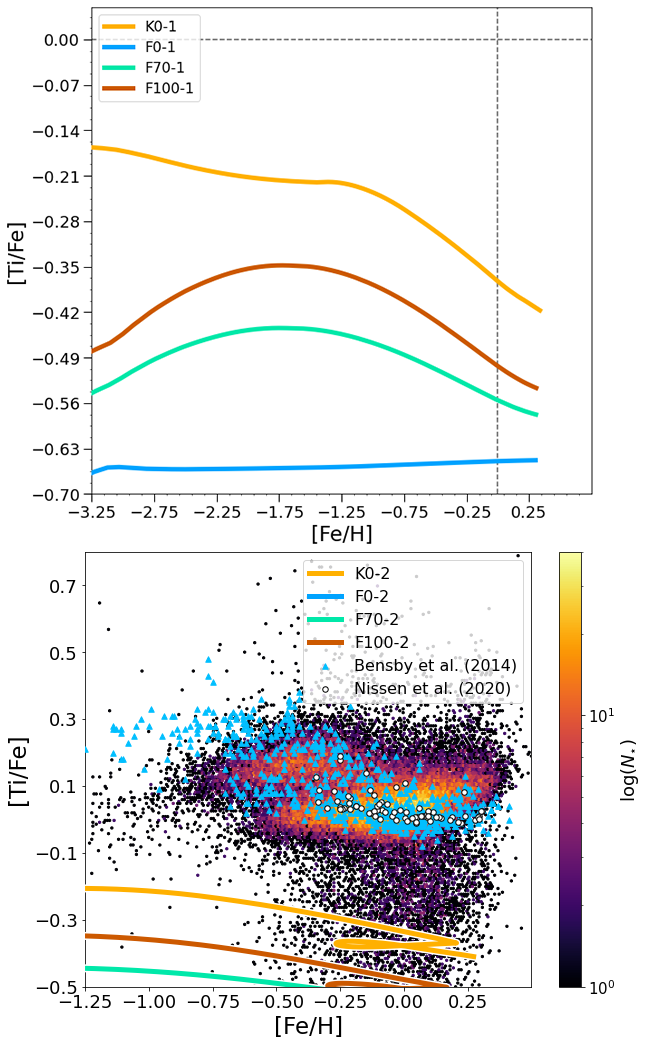}
    \caption{Same as Fig. \ref{C_map} but for [Ti/Fe]. Data are from \citet[azure triangles]{bensby2014}, \citet[white points]{nissen2020} and \citet[colored according to their number density, see colorbar]{apogeedr172022}.}
    \label{Ti_map}
\end{figure}

The last $\alpha$-element we examine is Ti, which is known to show critical issues on its nucleosynthetic production from different stellar evolution models, as pointed out by several authors (see, e.g. \citealt{romano2010,prantzos2018,kobayashi2020}). However, it is worth noting that the Ti production is crucially dependent upon the assumption of spherical symmetry in the generally adopted (1-dimensional) CC-SN models, and this can be overcome only by adopting multi-dimensional models \citep{rauscher2002, magkotsios2010, harris2017, sandoval2021}. As one can see, in spite of the yield differences, none of the  models adopted in this paper are able to reproduce the stellar abundances of Ti. In particular, the worst situation appears to be that of Model F0-2 with Farmer's yields for single massive stars. As said above, this large discrepancy could be solved by using multi-dimensional models for CC-SNe, but this is not the purpose of the present paper.

\subsubsection{Potassium}
\label{K_subsubsec}

Passing to odd-Z elements, we focus our attention on K, whose yields have been shown to severely underestimate the observed Galaxy abundance patterns (\citealt{romano2010,kobayashi2020} and references therein), even invoking mechanisms such as stellar rotation (\citealt{prantzos2018}).

\begin{figure}
    \centering
    \includegraphics[scale=0.390]{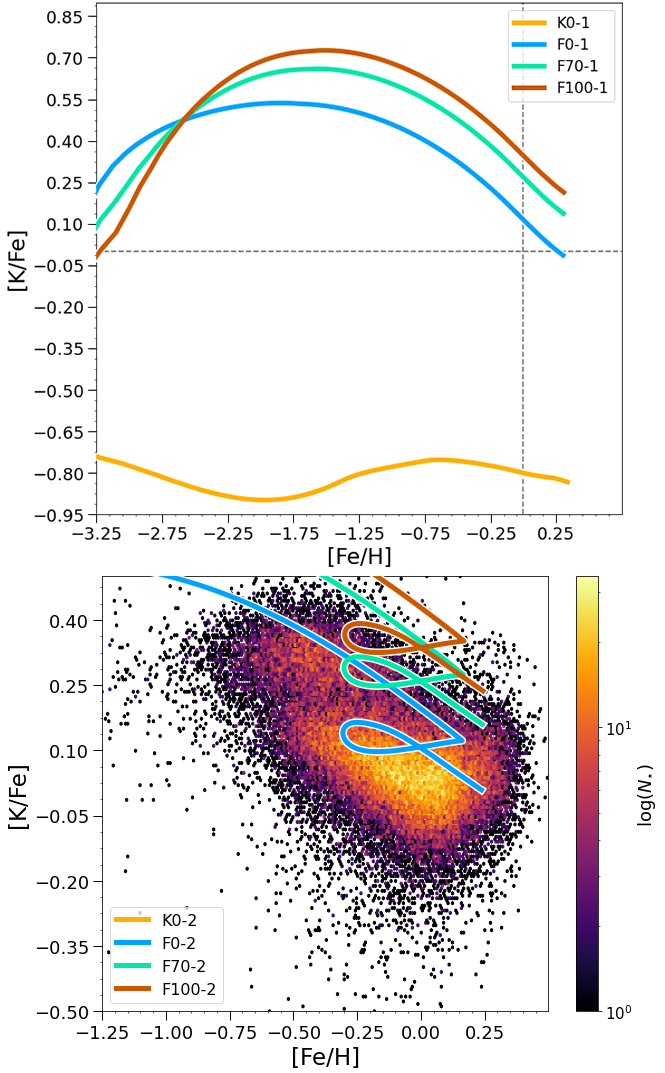}
    \caption{Same as Fig. \ref{C_map} but for [K/Fe]. Data are from \citet[colored according to their number density, see colorbar]{apogeedr172022}.}
    \label{K_map}
\end{figure}

In Figure \ref{K_map}, we thus compare results from our models for [K/Fe] vs. [Fe/H], testing the newly proposed massive star yields by \citet{Farmer2023}. 
In the upper panel,  the difference between the various sets of yields is striking. On the one hand, the Reference Model K0-1 shows a very low ratio of [K/Fe] at all metallicities, ranging from $\sim-0.6$ dex to $\sim-0.9$ dex. 
On the other side, the models adopting \citet{Farmer2023} yields show a very different pattern. 
From [Fe/H]$\sim-3.0$ dex towards higher metallicities, all models show an important increase in the [K/Fe] ratio, which depends on the fraction of binaries assumed in the chemical evolution model. 
The highest increase is observed for models with higher binary fraction, starting from sub-solar [K/Fe] ratio and reaching values higher than 0.6 dex before the decline caused by the Type Ia SN contribution. Assuming no binaries instead, the chemical evolution track starts from slightly supersolar [K/Fe] ratios and reaches values of the order of 0.5 dex. Such different behaviors are due to the fact that in \citet{Farmer2023} yield grids, very high mass stars in binaries are producing a lower amount of K, while lower  massive stars (which dominate the global contribution going towards larger metallicities) show more favorable conditions for K production relative to single stars in the same mass range. 
In fact, as suggested in \citet{Farmer2023}, potassium is greatly affected by the binary star presence, with the main contributors to its production being low mass binary-stripped stars during their pre-supernova evolution.

The radically different behavior between the Reference Model, representing what is generally predicted by the most used yield sets in the literature (\citealt{WW1995,koba2006,kobayashi2011,limongi2018}), and the models adopting \cite{Farmer2023} yields has important implications when comparing the results with the observed abundance patterns in the MW. 
This is shown in Figure \ref{K_map} lower panel:
while the Reference Model K0-2 falls even below the range of [K/Fe] values shown in the panel, the Model F0-2 (assuming \citealt{Farmer2023} yields without binaries) reproduces very well the observed abundance trend in the solar vicinity. It is worth noting that this is the first time in which chemical evolution models for the MW are reproducing the observed K trend without invoking ad-hoc assumptions in the yields (see for example \citealt{francois2004}). Concerning instead the models including massive binary-stripped star yields, they overestimate the observed [K/Fe] and this overestimation is proportional to the binary fraction.

\subsubsection{Chromium}
\label{Cr_subsubsec}

\begin{figure}
    \centering
    \includegraphics[scale=0.35]{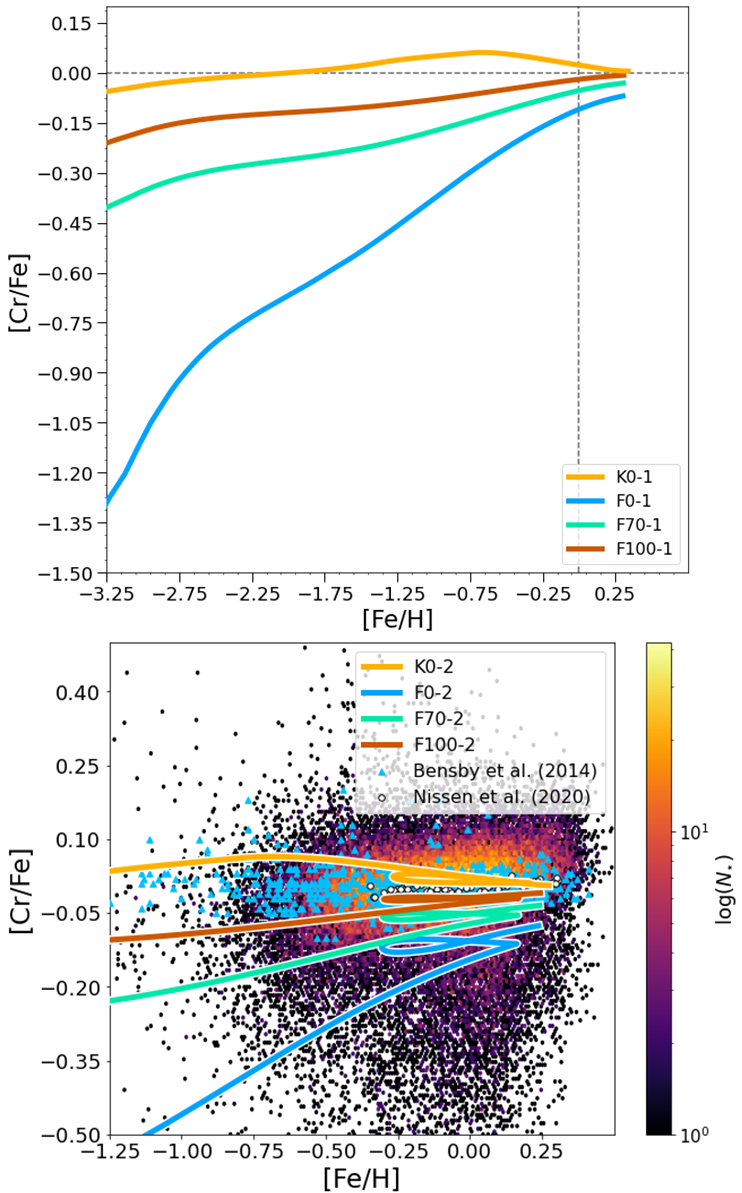}
    \caption{Same as Fig. \ref{O_map} but for [Cr/Fe]. Data from \citet[azure triangles]{bensby2014}, \citet[white points]{nissen2020} and \citet[colored according to their number density, see colorbar]{apogeedr172022}.}
    \label{Cr_map}
\end{figure}

For what concerns the Fe-peak group, the element showing the most significant variations between different yield sets is Cr. 
In Figure \ref{Cr_map} upper panel, we observe that the Reference Model K0-1 shows a rather flat, roughly solar pattern. The models adopting \citet{Farmer2023} yields with high binary fractions (F70-1 and F100-1), produce instead an increasing [Cr/Fe] ratio, with values starting from $-0.2$ dex (F100-1) and $-0.4$ dex (F70-1) and roughly reaching the solar ratio at [Fe/H]$\gtrsim$0.
The F0-1 Model, on the other hand, shows an extremely low [Cr/Fe] ratio when compared to the other models at low metallicity ([Cr/Fe] $\sim-1.4$ dex at [Fe/H] $\sim-3$ dex) 
and a progressive increase towards larger metallicities, reaching [Cr/Fe] $\sim -0.15$ dex at at [Fe/H]$\gtrsim$0. 
This indicates that in \citet{Farmer2023} stellar models, single massive stars are highly subdominant Cr producers relative Fe. This is at variance with binary-stripped stars where Cr is produced as the Fe, as found in most of the available literature prescriptions (see also \citealt{prantzos2018,koba2020b,palla2021}).

In Figure \ref{Cr_map} lower panel, 
we can see that the roughly flat behavior shown by the Reference Model and the \citet{Farmer2023} model with highest binary fraction (F100-2) well reproduce the trend of data from solar neighborhood stars presented in Section \ref{datasamp}.
The analysis of this panel excludes the Model F0-2 as a good tracer of Chromium in the solar vicinity, as it predicts a [Cr/Fe] increasing trend that is not observed either in APOGEE or \citet{bensby2014} and \citet{nissen2020} (see also \citealt{bergemann10,Adibekyan2012}).
For the remaining F70-2 Model, we see that its chemical tracks underestimate the observed Galactic Cr content as seen in \citet{bensby2014}, especially for metallicities below [Fe/H]$\lesssim -0.5$ dex.
In conclusion, both the Reference Model and the F100-2 Model reproduce the general trend. However, we cannot draw firm conclusions on which of the two is in better agreement with the observations. In fact, as the contribution of Type Ia SNe to Cr is comparable to the one by massive stars, different yield prescriptions for Type Ia SNe could change the chemical evolution picture (see \citealt{palla2021} and references therein) in favor of one or the other model.

\subsection{Discussion and caveats}
\label{caveats}

Throughout this work, we investigated the effects of yields from massive single and binary-stripped stars, as computed by \citet{Farmer2023}, on the chemical evolution of the Galaxy.
Previous work (e.g. \citealt{dedonder2002}) had already explored the effects of the yields from massive interacting binaries adopting a model similar to that of \citet{chiappini1997} and therefore similar to the model we adopted. The conclusions were that the inclusion of massive binaries in the Galactic chemical enrichment was not striking and the effects on the results were not larger than a factor of 2.
Recently, the work of \citet{Farmer2023} reopened the interest in studying the effects of chemical pollution from binary systems. The reason is that most of the massive stars in the Galaxy should reside in binary systems ($\gtrsim$70\%, e.g. \citealt{DeRosa14,Thies15}), as the vast majority of stars form as binaries and dynamical processes do not have the time to decrease significantly the binary fraction within the stellar lifetimes (\citealt{Kroupa24}).\\

Our analysis has shown that the adopted yields for massive stars in binaries can produce different solar abundances than yields adopting standard nucleosynthesis prescriptions from single massive stars (e.g. \citealt{romano2010} and references therein), and for some elements they lead to a better agreement with observations. 
However, a shortcoming of the \citet{Farmer2023} yield grids is that they are computed, either for stars in single or binary systems, for a unique stellar chemical composition, i.e. the solar one. On the other hand, most of the yields from single massive stars present in the literature are computed for different chemical compositions, from absent (e.g. \citealt{Heger10}) or very low (e.g. [Fe/H]=$-3$ dex, \citealt{limongi2018}) metal content up to the solar composition. 
 This fact can create large differences and a fair comparison will be possible only when the yields for stars in massive interacting binaries will be computed for different metallicities; this is particularly important for elements with prominent secondary component, such as $^{14}$N, the odd-Z elements $^{27}$Al and $^{23}$Na, $^{63}$Cu. 
In particular, the evolution of elements partly secondary, such as $^{14}$N, can be quite different, since the yields are very low at low metallicity and therefore the overall element production would result lower than assuming the yield computed for the solar metallicity during the whole chemical evolution, as it is done in our paper.

Another limitation of this work regards the fate and yields of secondary stars in binaries, which are treated here as single stars. This is actually inherited from the study of \citet{Farmer2023}, which does not provide the nucleosynthetic yields from such binary products. Their most significant contribution would likely be due to the mass gain, inducing stars that are initially too small to produce SNe (initial mass slightly lower than 8 M$_{\odot}$) to gain enough mass to explode (e.g. \citealt{2004podsiadlowski}; \citealt[\citeyear{2019Zapartas}]{2017Zapartas}). However, the picture is even more complex since their contribution to the Galactic chemical enrichment also depends on: i) the fraction of mass that they accreted and remained bound and ii) the amount of such accreted bound mass, further processed inside the secondary star (see e.g. \citealt{2021Deckers}).

Finally, another source of uncertainty in the yield computation is the adopted stellar evolution code. Different codes include different treatments of the physical processes (e.g. convection, SN induced explosion) inside the stars and therefore produce different results.\\

Here, therefore, we do not intend to draw firm conclusions on the effects of binaries on the chemical enrichment process, neither establish the right percentage of massive binaries, as we also  lack in well sampled grids of binary system parameters such as mass ratios and orbital periods, restricting to those in which all massive binaries undergo case B mass transfer, producing binary-stripped primary stars (see \citealt{Moe17} for a review on binary parameter distribution).
Rather, the goal of this paper is to indicate that the inclusion of yields from massive stars in binaries can affect the agreement between model results and observed abundances.

In this way, our study wants to both i) encourage for further development in stellar yield modeling in binary systems enlarging the grids in metallicities and binary systems conditions, and ii) provide a first base for future studies which will investigate the influence of binary systems also from the point of view of Galactic chemical evolution.

\section{Conclusions}
\label{conclu_sec}

In this study, we have compared the results from detailed and well-tested chemical evolution models for the MW (e.g. \citealt{spitoni2019,palla2020}) including either yields from single massive stars or yields considering the contribution of massive stars in interacting binaries. 

In particular, we tested the yields from massive binary-stripped stars as computed by \citet{Farmer2023} and assumed different binary fractions in the IMF from 0\% to 100\%. 
We also consider standard yields for single massive stars (\citealt{romano2010}, i.e. our Reference Model) largely adopted in previous chemical evolution works, to facilitate the comparison with what is obtained with available literature prescriptions.
Our goal was to test whether the inclusion of yields of massive stars in binaries could substantially change the modeled abundance patterns in the solar vicinity. 
It is worth noting that such results are relevant, as in the last two decades no other studies addressed the effects of massive binary systems specifically on the chemical evolution of the Galaxy ( \citealt{dedonder2002}).\\

Our key findings can be summarized as follows:
\begin{enumerate}
    \item when adopting \citet{Farmer2023} stellar yields the differences in the predictions obtained by varying the percentage of binary stars are negligible for a large fraction of the studied chemical elements;
    
    \item a more marked difference is instead found for both the predicted solar abundances and the abundance patterns ([X/Fe] vs. [Fe/H]) between the Reference Model (K0-2), adopting standard yields for single massive stars dependent on metallicity, and models adopting \citet{Farmer2023} stellar yields for single massive stars at solar metallicity;

    \item by adopting \citet{Farmer2023} yields for single stars (Model F0-2), the solar abundances predicted by the chemical evolution models are more in agreement with observations relative to the ones predicted by the Reference Model K0-2 for $^{4}$He, $^{12}$C, $^{14}$N, $^{24}$Mg, $^{39}$K, $^{40}$Ca, $^{55}$Mn and $^{59}$Co. 
    Of these elements, when considering [X/Fe] vs. [Fe/H] diagrams, the [C/Fe] ratio as predicted by F0-2 Model presents some difficulties in reproducing all the data, although it is in relatively good agreement with \citet{nissen2020} and \citet{apogeedr172022} data for solar and supersolar metallicities. The [Mg/Fe] predicted ratio shows instead a remarkable agreement with both the high-$\alpha$ and low-$\alpha$ sequence, which is not found with the yields adopted in the Reference Model, as well as for other yield prescriptions available in the literature (see \citealt{palla2022}).
    Also the predicted [Ca/Fe] vs. [Fe/H] relation very well reproduces the observational data from all samples, especially at high metallicity. 
    Lastly, the Model F0-2 is able to reproduce the  observed $^{39}$K trend without invoking ad-hoc assumptions for the yields, as instead required by most of the available stellar yield prescriptions (e.g. \citealt{kobayashi2020});

    \item regarding models that use the \citet{Farmer2023} yields with different fractions of binary systems, when comparing their predicted solar abundances to the Reference Model K0-2, we find that  Model F70-2 (70\% binary fraction) shows better agreement for $^{24}$Mg, $^{40}$Ca, and $^{56}$Fe, while F100-2 shows better results for  $^{40}$Ca, $^{52}$Cr and $^{56}$Fe than the Reference Model. 
    For both [Ca/Fe] and [Mg/Fe] the inclusion of binary-stripped yields with different fractions produce similar pattern to that of \citet{Farmer2023} yields with single stars only, thus allowing a good reproduction of the observed abundance patterns in the solar vicinity.
    Finally, [Cr/Fe] ratio is best reproduced by F100-2, whereas F70-2 slightly underestimates the trend observed in \citet{bensby2014};

    \item both the $^{16}$O predicted solar abundances and [O/Fe] vs. [Fe/H] relation, are generally overestimated by all models, and especially by F0-2 Model, which does not reproduce at all the observed data trend.
    On the other hand, the solar abundances and abundance patterns predicted by all models for $^{48}$Ti are largely underestimated, with the results of Model F0-2 increasing significantly the already present discrepancy with the observed [Ti/Fe] data trend;
    
    \item for the other chemical elements, we do not observe noticeable differences in the predicted solar abundances as well as in the [X/Fe] vs. [Fe/H] ratios when comparing the results obtained by means of different yields. We also avoid to draw conclusions on the abundance patterns of elements with relevant secondary production (e.g. $^{14}$N), as the new yields of \citet{Farmer2023} are currently computed only for solar metallicity, thus preventing a fair comparison with other models and observations.
    
\end{enumerate}

\begin{acknowledgements}
 The authors want to thank the anonymous referee for the important and useful suggestions improving the manuscript content.
MP acknowledges financial support from the project "LEGO – Reconstructing the building blocks of the Galaxy by chemical tagging" granted by the Italian MUR through contract PRIN2022LLP8TK\_001.
F. Matteucci thanks I.N.A.F. for the 1.05.12.06.05 Theory Grant - Galactic archaeology with radioactive and stable nuclei.    F. Matteucci thanks also support from Project PRIN MUR 2022 (code 2022ARWP9C) “Early Formation and Evolution of Bulge and HalO (EFEBHO)” (PI: M. Marconi). E. Spitoni  thanks I.N.A.F. for the  1.05.23.01.09 Large Grant - Beyond metallicity: Exploiting the full POtential of CHemical elements (EPOCH) (ref. Laura Magrini).
In this work, we have made use of SDSS-IV APOGEE-2 DR17 data. Funding for the Sloan Digital Sky Survey IV has been provided by the Alfred P. Sloan Foundation, the U.S. Department of Energy Office of Science, and the Participating Institutions. SDSS-IV acknowledges support and resources from the Center for High-Performance Computing at the University of Utah. The SDSS web site is  \href{www.sdss.org}{www.sdss.org}.
SDSS is managed by the Astrophysical Research Consortium for the Participating Institutions of the SDSS Collaboration which are listed at \href{https://www.sdss.org/collaboration/affiliations/}{www.sdss.org/collaboration/affiliations/}. 
\end{acknowledgements}
\bibliographystyle{aa} 
\bibliography{pepe}

\end{document}